\documentclass{emulateapj}

\slugcomment{accepted to ApJ}
\shorttitle{Environmental Effects}
\shortauthors{Khochfar \& Ostriker}

\begin{document}
\title{Adding Environmental Gas Physics to the Semi-Analytic Method for 
Galaxy Formation: Gravitational Heating}
\author{Sadegh Khochfar }
\affil{ Department of Physics, University of Oxford}
\affil{Denys Wilkinson Bldg., Keble Road, Oxford, OX1 3RH, UK}
\email{sadeghk@astro.ox.ac.uk}
\and
\author{Jeremiah P. Ostriker}
\affil{Department of Astrophysical Sciences, Princeton University}
\affil{Peyton Hall, Princeton, NJ 08544, USA}
\email{ostriker@princeton.edu}

\begin{abstract}
We present results of an attempt to include more detailed gas physics motivated 
from hydrodynamical simulations within 
semi-analytic models (SAM) of galaxy formation, focusing on the role that 
environmental effects play. The main difference to previous SAMs is that we include
 'gravitational' heating of the intra-cluster medium (ICM) 
by the net surplus of gravitational potential energy released from gas that 
has been stripped from infalling satellites. Gravitational heating appears 
to be an efficient heating source able to prevent cooling in environments
 corresponding to dark matter halos more massive than  $\sim 10^{13} $M$_{\odot}$.
The energy release by gravitational heating  can match that by AGN-feedback 
in  massive 
galaxies and can exceed it in the most massive ones. 
However, there is a fundamental difference in the way the two processes operate. 
Gravitational heating 
becomes important at late times, when the peak activity of AGNs is already over, and 
it is very mass 
dependent. This mass dependency and  time behaviour  gives the right trend to 
recover down-sizing in the star-formation rate of massive 
galaxies. In general we find that 
environmental effects play the largest role in halos more massive than $M^*$ at any 
given redshift because of the continued growth by mergers in these halos.
We present a number 
of first order comparisons of our model to well established observations 
of galaxy properties which can be summarised as follows: The cosmic star formation 
rate can be reproduced with a decline at $z < 1$ that is steeper with respect to 
our SAM without environmental effects. The steep decline is 
mainly driven by a suppression of star formation in high density environments. 
In addition the star formation episode of our model galaxies is a strong function 
of mass. Massive galaxies with M$_* > 10^{11}$ M$_{\odot}$ 
make most of their stars at look back times of roughly 11 Gyrs and 
show very low amounts of residual star formation at 
late times due to the suppression by environmental effects. 
In addition the luminosity function and  colour bi-modality of the galaxy population 
are reproduced well.

\end{abstract}


\section{Introduction}
The simplest semi-analytic model (SAM) for galaxy formation 
might assume that the nature of a galaxy is 
determined solely by the mass and merging history of the dark matter halo or 
sub-halo within which it resides.
But correlations between the observed properties of galaxies and 
their environments have been 
known for many years now. An increased fraction of early-type galaxies towards the 
centres of clusters \citep{1980ApJ...236..351D,1982ApJ...263..533D,2005ApJ...620...78S}, 
or the increased fraction of blue, star forming galaxies
 in clusters at high redshift compared to their low redshift counter parts, the 
Butcher-Oemler effect \citep{1978ApJ...219...18B,1984ApJ...285..426B}. 
Massive elliptical galaxies in clusters are on  
average older than comparable ellipticals in the field 
\citep{2005ApJ...621..673T} and  the interaction rate of galaxies in clusters increases with 
redshift much more rapidly than in the field \citep{1999ApJ...520L..95V} indicating 
accelerated galaxy transformation in dense environments. Recent studies of e.g. the 
Sloan Digital Sky Survey (SDSS) and the DEEP2 survey show that the fraction of galaxies 
of a given mass populating the red-sequence is a strong function of the environment 
\citep{2006MNRAS.373..469B} and that indeed the environment can play an import role 
\citep{2006MNRAS.370..198C} in driving galaxy evolution.    

On the theoretical side first indictations for a possible transition in the 
mode of galaxy formation were found by  \citet{1977MNRAS.179..541R}, \citet{1977ApJ...211..638S} and \citet{1977ApJ...215..483B} 
who predicted a transition mass scale at which 
cooling times start becoming larger than the free-fall time. 
Differences arising from the environment are generally 
studied by assuming that the dark halo mass is a good proxy for the environment and that 
dark matter haloes greater than a few times  $10^{14}$ M$_{\odot}$ correspond 
to galaxy clusters (for a comparison between dark halo mass and 
galaxy surface density see \citet{2006MNRAS.373..469B}). Regarding the formation of 
dark matter halos and the mass assembly of galaxies, the CDM-paradigm  predicts
massive dark haloes to have a larger fraction of their final mass in progenitors at 
earlier times compared to low mass halos \citep[e.g.][]{2006MNRAS.372..933N}
resulting in an increased merger fraction of galaxies at larger redshifts in 
high density environments \citep{2001ApJ...561..517K}. Furthermore, galaxies 
falling into dense environments are subject to interaction between the 
hot intra-cluster medium (ICM) and the inter-stellar medium (ISM) in form of 
ram-pressure stripping 
\citep{1972ApJ...176....1G,1980ApJ...241..928F,1999MNRAS.308..947A,2000ApJ...538..559M}
 and shock heating \citep[e.g.][]{1999ApJ...525..554F} that can allow for morphological 
transformation of galaxies by depriving them of gas \citep{2000Sci...288.1617Q}
  and inducing star formation \citep{2007ApJ...654..825M}. The interplay between 
these processes will influence galaxies within a dark matter halo and should  
in a natural way account for many observations. Models based on a 
phenomenological approach to stop cooling flows in halos above a critical mass  
\citep[e.g][]{1999MNRAS.303..188K,2004MNRAS.347.1093B}
and hence in dense environments  prove to be promising by being able to reproduce the 
luminosity function and colour bi-modality of the local galaxy population 
\citep[e.g][]{2006MNRAS.370.1651C}. Many of these effects are 
already seen in hydrodynamic simulations \citep{1999ApJ...522..590B,2000ApJ...531....1B}.

The purpose of this paper is to illustrate the consequences that result from 
environmental effects usually omitted 
 within semi-analytic models (SAM). These effects
are real and automatically included in  hydrodynamical simulations of sufficient resolution 
\citep[see e.g.][]{2005astro.ph.12235N}, but we do not
intend to claim that the implementation attempted in this paper is definitive or even,
necessarily a substantial improvement over current modeling methods. Rather we 
intend to show the  sign of the effects and the rough order of magnitude of the 
effects produced. In many cases these physical effects produce consequences that 
seem at variance (or even in some cases opposite) to overly naive interpretations 
of the hierarchical scenario that are consequent to gravitational interactions alone. 
Also, we wish to make clear at the outset that the physical effects described in 
this paper are independent of "feedback" which concept we define to be related to 
the return of energy, momentum and mass to the environment from evolving stars and 
central black holes. These feedback effects are also real, but they should not be
confused with those gas physics effects that would necessarily occur even if there 
were neither stars nor AGNs. If the effects that we are adding to the SAM 
prescription are found to be useful, the correct way to implement them will be to 
use comparisons with detailed high resolution numerical hydro simulations, which 
are just now becoming available, and of course feedback effects from central 
black holes 
\citep{2001ApJ...551..131C,2005ApJ...620L..79S,2006MNRAS.365...11C,2006MNRAS.370..645B,2006ApJ...648..820K,2007astro.ph..3057C} 
must also be added to a comprehensive treatment for most accurate results.

We briefly summarise here the additional gas physics we have 
included to our basic model introduced in section 2 that made the most 
important differences in the model predictions, before we show later 
the details of the implementation (section 3) and first results (sections 4-8).
For individual satellite galaxies we 
relaxed the generally applied assumption of instantaneous shock heating of hot gas 
in them when they fall into dense environments (Section \ref{sshock})  
and added a prescription for ram-pressure stripping of gas (Section \ref{sram}) 
while they orbit in a dense environment. We modify cooling flows for central 
galaxies, generally the most massive galaxies within a dark matter halo, by taking 
into account the heating of the ICM by  gravitational heating from potential energy 
released by stripped gas from satellites (section \ref{grav}). 
This additional energy source, which is 
not usually included in SAM treatments (but automatically allowed for in hydro 
treatments) adds energy to the gas in amounts comparable to the energy added by 
feedback processes.

\section{The Semi-analytic model}
The main strategy behind the modelling approach we follow is first to calculate
 the collapse and merging history of individual dark matter halos, 
which is governed purely by  gravitational interactions, and secondly 
to estimate the more complex physics of the baryons inside 
these dark matter halos, including  e.g. radiative cooling of the gas, star formation, 
and feedback from supernovae by simplified prescriptions on top of 
the dark matter evolution \citep[e.g.][]{1977ApJ...211..638S,1978MNRAS.183..341W,1991ApJ...379...52W,1999MNRAS.303..188K,1999MNRAS.310.1087S,2000MNRAS.319..168C,2003MNRAS.343...75H,2006MNRAS.370.1651C,2006MNRAS.365...11C,2006MNRAS.370..645B,2007MNRAS.375....2D}. Each of the dark matter halos will consist of three 
main components which are distributed among individual galaxies inside  
them: a stellar, a cold, and  a hot gas component, where the latter  is
only attributed to {\it central} galaxies, which are the most massive galaxies
inside individual halos and typically are observed to reside in extended X-ray 
emitting coronal gas.
In the following sections, we will describe briefly the recipes used to
 calculate these different components which 
are mainly based on recipes presented in e.g. 
\citet{1999MNRAS.303..188K} (hereafter, K99), \citet{2000MNRAS.319..168C} (hereafter, C00) and 
\citet{spr01} (hereafter, S01), and we refer readers for more details 
on the basic model implementations to their work and references therein. 
In the remainder of this paper we call the {\it standard/old} SAM the model implementation 
presented in \citet{2003ApJ...597L.117K,2005MNRAS.359.1379K}  which is summarized in 
this section. The new additional environmental physics is presented in section \ref{envir}. 
Throughout this paper we use the following set of cosmological 
parameters motivated by the 3 year
results of WMAP \citep{2006astro.ph..3449S}:
$\Omega_0=0.27$, $\Omega_{\Lambda}=0.73$, $\Omega_b/\Omega_0=0.17$, 
$\sigma_8=0.77$ and $h=0.71$.

\subsection{Dark Matter Evolution}
We calculate the merging history of dark matter halos according to 
the prescription presented in \citet{1999MNRAS.305....1S}. This approach has 
been shown to produce merging histories and progenitor distributions in 
reasonable agreement with results from N-body simulations of cold dark matter 
structure formation in a cosmological context \citep{2000MNRAS.316..479S}. 
The merging history of dark matter halos is reconstructed by breaking 
 each halo up into progenitors above a limiting minimum progenitor 
mass $M_{min}$. This mass cut needs 
to be chosen carefully as it ensures that the right galaxy population and 
merging histories are produced within the model. Progenitor halos
 with masses below $M_{min}$ are declared as {\it accretion} events and 
their histories are not followed further back in time. 
Progenitors labelled as accretion events should ideally not host any 
significant galaxies in them and be composed mainly of primordial hot gas 
at the progenitor halo's  virial temperature. 
The mass scale at which this is the case can in principle be 
estimated from the prescriptions of supernova feedback and reionisation 
presented in section \ref{cool}.
However,  to achieve a 
good compromise between accuracy and computational time, we instead estimated 
$M_{min}$  by running several simulations with different resolutions 
and chose the resolution for which results in the galaxy mass range of 
interest are independent of the specific choice of $M_{min}$. 
Changing the mass resolution mainly affects our results at low galaxy mass 
scales, leaving massive galaxies nearly unaffected.
Throughout this paper we will use $M_{min}= 10^{10}$ M$_{\odot}$  
which produces numerically stable results for galaxies with stellar masses 
greater a few times $10^{10}$ M$_{\odot}$.

\subsection{Baryonic Physics}
As mentioned above, once the merging history of the dark matter 
component has been calculated, it is
 possible to follow the evolution of the baryonic content in these halos 
forward in time. We assume each halo  consists of three components: 
hot gas, cold gas and stars, where the latter two components can be distributed 
 among individual galaxies inside a single  dark matter halo. The stellar components 
of each galaxy are additionally divided into bulge and disc, to allow 
morphological classifications of model galaxies. In the following, we 
describe how the evolution of each component is calculated. 

\subsubsection{Gas Cooling \& Reionisation}\label{cool}
Each branch of the merger tree starts at a progenitor mass of $M_{min}$ and 
ends at a redshift of $z=0$. Initially, each halo is occupied by hot 
primordial gas which was captured in the potential well of the halo and shock 
heated to its virial temperature 
$T_{vir}=35.9\left[V_c/(\mbox{km s}^{-1}) \right]^2$ K, where $V_c$ is the 
circular velocity of the halo \citep[K99]{1991ApJ...379...52W}. Subsequently this hot 
gas component is allowed to radiatively cool and settles down into a
rotationally supported gas disc at the centre of the halo, which we identify 
as the central galaxy \citep[e.g][]{1977ApJ...211..638S,1978MNRAS.183..341W,1991ApJ...379...52W}.  
The rate at which hot gas cools down is estimated by calculating the 
cooling radius inside the halo using the cooling functions provided by 
\citet{1993ApJS...88..253S} and the prescription in S01. In the case of a merger 
between halos we assume that all of the hot gas present in the progenitors 
is shock heated to the virial temperature of the remnant halo (in our new model including 
environmental effects we will relax this assumption, see \ref{sshock}), and that gas 
can only cool down onto the new central galaxy which is the central galaxy 
of the most massive progenitor halo. The central galaxy of the less massive 
halo will become a satellite galaxy orbiting inside the remnant halo. In this 
way, a halo can host multiple satellite galaxies, depending on the 
merging history of the halo, but will always only host one central galaxy onto 
which gas can cool. The cold gas content in  satellite galaxies  is given by 
the amount present when they first became satellite galaxies and does 
not increase (this will again will be modified in our new environmental prescription, 
see \ref{sshock}, instead it decreases due to ongoing star formation and supernova 
feedback.

In the simplified picture adopted above, the amount of gas available to 
cool down is only limited by the universal baryon fraction 
$\Omega_b h^2=0.023$ \citep{2006astro.ph..3449S}. However, in the presence of a 
photoionising background the fraction of baryons captured in halos is reduced
\citep[e.g.][]{2000ApJ...542..535G,2002MNRAS.333..156B} and we use the recipe of \citet{2002ApJ...572L..23S}, 
which is based on a fitting formulae derived from hydrodynamical simulations 
by \citet{2000ApJ...542..535G}, to estimate the amount of baryons in each halo. For the epoch 
of  reionisation, we assume $z_{reion}=7$, which is in agreement with 
observations of  
the temperature-polarisation correlation of the cosmic microwave background 
by \citet{2006astro.ph..3449S}. 

\subsubsection{Star formation in Discs and Supernova Feedback}\label{sf}
Once cooled gas has settled down in a disc, we allow for fragmentation and 
subsequent star formation according to a parameterised global 
Schmidt-Kennicutt type law  \citep{1998ApJ...498..541K} of the form 
$ \dot{M}_{*}=\alpha M_{cold}/t_{dyn,gal}$, where $\alpha$ is a free parameter
describing the efficiency of the conversion of cold gas into stars, and 
$t_{dyn,gal}$ is assumed to be the dynamical time of the galaxy and is 
approximated to be 1/40 times the dynamical time of the dark matter halo 
\citep{2006MNRAS.366..899N}.

Feedback from supernovae plays an important role in regulating star 
formation in small mass halos and in preventing too massive satellite 
galaxies from forming \citep{1986ApJ...303...39D}. 
We implement feedback based on the prescription presented in K99 with
\begin{equation}
  \Delta M_{reheat}=\frac{4}{3} \epsilon \frac{\eta_{SN} E_{SN}}{V_{c}^{2}} 
  \Delta M_*. 
\end{equation}
Here we introduce a second free parameter $\epsilon$ which represents our 
lack of knowledge on  the efficiency with which the energy from supernovae 
reheats the cold gas.  The expected number of supernovae 
per solar mass of stars formed for a typical IMF is  $\eta_{SN}=5 \times 10^{-3}$,  
 and $E_{SN}=10^{51}$ erg  is the energy output from each supernova. We take 
$V_c$ as the circular velocity of the halo in which the galaxy was  
last present as  a central galaxy.

\subsubsection{Galaxy Mergers}\label{merger}
We allow for mergers between galaxies residing in a single halo. As mentioned 
earlier, each halo is occupied by one central galaxy and a number of 
satellite galaxies depending on the past merging history of the halo. 
Whenever two halos merge, the galaxies inside  them will merge on a 
time-scale which we calculate by estimating the time it would take the 
satellite to reach the centre of the halo under the effects of 
dynamical friction. Satellites are assumed to merge only with 
central galaxies and we set up their orbits in the halo according to the 
prescription of K99, modified to use the Coulomb logarithm 
approximation of S01.

If the mass ratio between the two merging galaxies is $M_{gal,1}/M_{gal,2} 
\leq 3.5$ ($M_{gal,1} \geq M_{gal,2}$) we declare the event as a 
{\it major} merger and the remnant will be an elliptical galaxy. We assume that 
the stellar components of 
the progenitors add up to form a spheroid, that 
the cold gas present in the progenitors ignites in a central star burst, and that  
the hot gas components add up. In the case 
of a {\it minor} merger $M_{gal,1}/M_{gal,2} > 3.5$, 
we assume that the stellar component of the smaller progenitor adds to the bulge of 
the larger progenitor and that the cold gas in the disc of the smaler 
progenitor settles down into the disc of the larger progenitor. 

\section{Environmental Effects}\label{envir}
The basic model introduced in the previous section
already includes several environmental
dependencies as e.g. the merger rate of galaxies increase more steeply with
redshift in high density environments like clusters 
\citep{2001ApJ...561..517K}, a feature also seen in observations 
 \citep{1999ApJ...520L..95V}. In this section we intend to incorporate
 further physical effects that are important to model the evolution of the 
galaxy population and that are already self-consistently included in 
hydrodynamical simulations \citep[e.g.][]{1999ApJ...525..554F}. Of these effects some 
were already implemented in similar ways in previous models by other authors, like e.g. 
ram-pressure stripping and shock heating \citep[e.g.][]{2005MNRAS.361..369L}. Some of the effects 
of gravitationl heating have been discussed recently by \citet{2007astro.ph..1363W} 
but have not yet been implemented within the context of a SAM.

The main difference to the SAM implementation of the previous section is that we here 
follow explicitly the heating of the hot gas phase by the conversion of gravitational potential energy.
To avoid adding liberated energy twice to the hot gas phase we adopt the following prescription. 
Initially, when the dark matter halo becomes more massive than $M_{min}$, we set the temperature of 
the hot gas to $T_{vir}$. During the subsequent growth of the dark matter halo 
however, we do not automatically increase the temperature of the hot gas in the 
host to the new virial temperature of the host dark matter halo (by host dark matter halo we 
always refer to the dark matter halo mass including all sub-structure).
 Instead  we let it only increase by the energy 
gained from the potential energy of the gas that is stripped from the substructure 
(see parargraph \ref{grav} and \ref{impl}).
 At each step in our simulation we keep track of the specific energy of the hot gas 
component and thus are able to calculate its temperature $T_{gas}$. 
This allows us to define the parameter $f_{e}=1+(T_{vir}-T_{gas})/T_{vir}$ to calculate the 
energy needed to remove gas from a halo. We here assume that the cold gas phase has $T_{gas}=0$
which sets $f_e=2$ for cold gas and $f_e=1$ for gas at the virial temperature 
of the halo. In general the liberated potential energy is sufficient to raise the temperature of 
the hot gas to the new virial temprature on a short time scale.
Another main difference to the old SAM is that we allow satellites to have hot gas which 
is able to cool  following the prescription in \ref{cool} and to subsequently form stars in the 
disk of the satellite galaxy. We remove hot gas from the satellites using the prescriptions 
laid out in sections \ref{sshock} \& \ref{sram}.

The following sections are structured as follows, first we introduce 
our prescriptions for the stripping of gas from orbiting satellites by ram pressure 
and shock heating. This is necessary to help
us approximate the rate at which potential energy is relased at each individual 
time step within our simulation. In the following two sections  we then introduce 
the actual prescriptions for gravitational heating and its implemetation within our SAM.

\subsection{Shock-heating of Gas}\label{sshock}
During the infall of satellite galaxies into dense environments, 
shock heating of satellite gas  is occurring   with the gas being
 removed from the satellite on time scales much shorter than the Hubble 
time \citep[e.g.][]{1994ApJ...437..564M,1999ApJ...525..554F}. 
Generally this process is implemented within SAMs by
assuming that it is very efficient and  quasi-instantaneous, thus depriving 
satellite galaxies of any reservoir of hot virialized gas 
(e.g. WF91, K99, C00). In the following we will relax this assumption and
investigate its effects on the galaxy population.

\begin{figure}
    \plotone{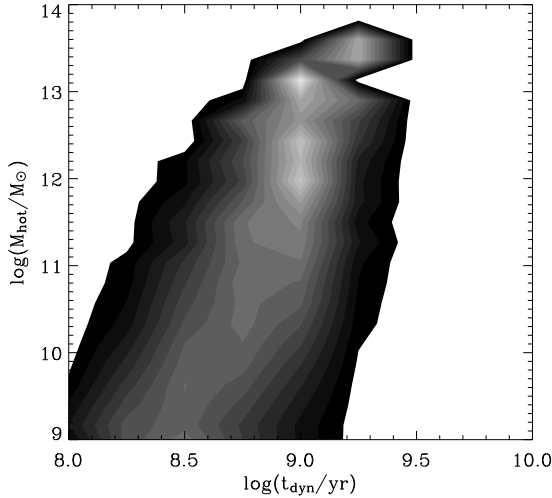} 
    \figcaption{Conditional probability contours $p(t_{dyn}|M_{hot})$ of 
      satellite galaxies when they enter the host dark matter halo. 
      The dynamical time $t_{dyn}$ has been calculated for the host 
      dark matter halo and $M_{hot}$ is the amount of hot gas within the 
      satellite galaxy. \label{fig2}}  
\end{figure}

Gas removal from the satellite occurs efficiently when
the shock is depositing enough energy to heat the gas in the satellite above 
its virial temperature \citep[e.g][]{1994ApJ...437..564M,2003MNRAS.345..349B}. 
This condition can be expressed in terms of the adiabatic sound
speed in the host halo, $c^2_{\mbox{gas}}=\gamma V^2_{max}/ \beta$ with 
$\gamma=5/3$ and $\beta\sim 1.25$ \citep{1998ApJ...495...80B}, and 
the satellite's maximum circular velocity $ V_{max,sat}$ as 
\begin{equation}\label{shock}
  c^2_{\mbox{gas}} \geq \zeta V^2_{\mbox{max,sat}}.
\end{equation}
Here we introduce a free model parameter $\zeta$ which in 
effect regulates at which mass ratio, between infalling satellite and 
host halo, shock heating will occur. 
As we will show below the specific choice of $\zeta$ does not change the  properties of 
the overall galaxy population significantly and only has influence on satellite galaxies.
The gas fraction of galaxies within clusters
is an increasing function of radius \citep{1986ApJ...301...35D} supporting the assumption that 
whatever reduces the gas fraction must work on a time scale comparable 
to the dynamical time within the cluster. Once the condition in Eq. \ref{shock} is 
satisfied, we therefore allow for gas removal from the satellite 
on some fraction $1/\delta$ of the host halo's dynamical time. In general 
we find that the gas in our simulations will be completely shock 
heated on time scales less than a Gyr with a tendency for faster 
heating in low mass satellites as shown by the conditional probability 
distribution $p(t_{dyn}|M_{hot})$ of infalling satellite galaxies
in Fig. \ref{fig2}. 

\begin{figure}
    \plotone{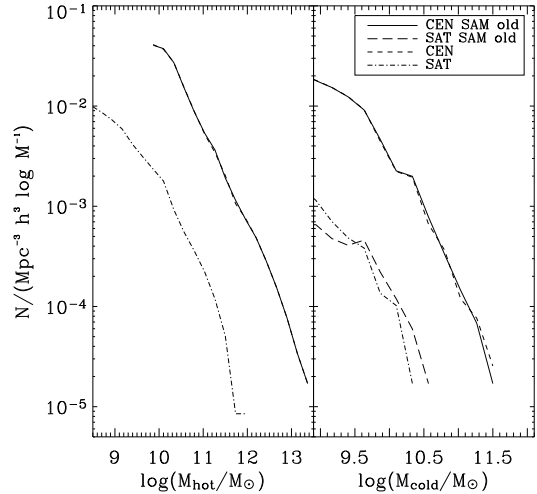} 
    \figcaption{Hot and cold gas mass function for satellite (SAT) and 
      central galaxies (CEN) in the model including shock heating of hot 
      and cold satellite gas. The initial semi-analytic model without 
      environmental effects is labelled 'SAM old'. Please note that the initial
      SAM by construction does not have any satellite galaxies with hot gas 
      reservoirs because we assume instantaneous heating of it. \label{fig5}}  
\end{figure}

\begin{figure}
    \plotone{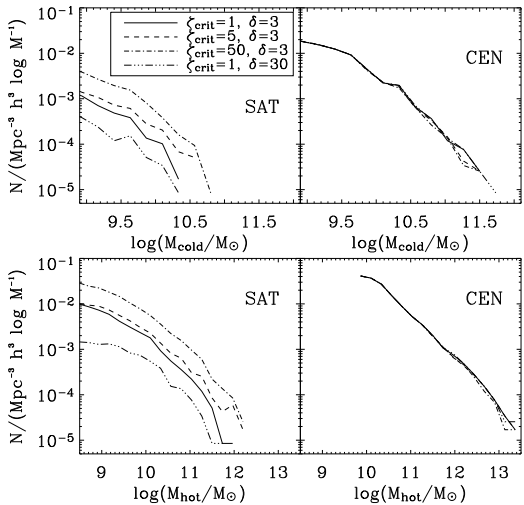} 
    \figcaption{Hot and cold gas mass functions for different choices of 
      parameters in the shock heating model. 
      We omit showing the initial 
      SAM because it is almost identical to the model with $\delta=3$ and 
      $\zeta=1$. \label{fig7}}  
\end{figure}

Shocks will  not only heat  the hot gas phase, but  also 
penetrate deep within the satellite and heat the cold gas phase (T. Naab, private communication).
We include this effect, sometimes neglected within SAMs, in the same manner as 
described above. In consequence at each time step within the simulation we 
have the following mass loss from each satellite due to shock heating which will 
be added to the hot gas content of the host central galaxy:
\begin{eqnarray}\label{shock2}
 \frac{dM_{\mbox{hot,sat}}}{dt}&=&-\delta\frac{M_{\mbox{hot,sat}}}{t_{dyn,halo}}\\
 \frac{dM_{\mbox{cold,sat}}}{dt}&=&-\delta\frac{M_{\mbox{cold,sat}}}{t_{dyn,halo}}.
\end{eqnarray}
A first comparison between our initial model (labelled 'old SAM'),
including  instantaneous shock heating of only hot gas, in Fig.
 \ref{fig5} does not show any significant difference in the cold and 
hot gas mass function of central galaxies. On the contrary Fig. \ref{fig7} 
shows that satellites tend to 
retain more cold and hot gas when the condition for shock heating, as laid out 
by Eq. \ref{shock}, requires them to fall into a much more massive 
host (larger $\zeta$) and if the time scale for shock heating is longer
(smaller $\delta$). In all cases the amount of gas shock heated is negligible 
as compared to the hot gas content of the host central galaxy, explaining the 
lack of change in the gas mass function. The overall galaxy population at the
intermediate to massive end is dominated by central galaxies which are 
unaffected by the specific choices of the shock heating parameters and we 
choose to omit these parameters for the remainder of this paper, i.e. set them to 
$\zeta=1$ and $\delta=1$.

\subsection{Ram Pressure Stripping}\label{sram}
Individual late-type galaxies within clusters show perturbed HI disks 
which are reduced in size with respect to their stellar disks 
\citep[for a review see][]{2004cgpc.symp..305V} and are HI deficient with 
respect to field late type galaxies, a fact generally  attributed to ram 
pressure stripping caused by the interactions between the hot ICM and the
 ISM of the satellite \citep{1972ApJ...176....1G,1983AJ.....88..881G,
2000Sci...288.1617Q}.
The time scale for this process must be less than 1 Gyr as indicated by the abrupt truncation of 
star formation in passive cluster spirals \citep{2006ApJ...641L..97M}. However, it is very likely 
that ram pressure stripping alone is not sufficient to cause the observed HI deficit in satellite 
galaxies \citet{1999MNRAS.308..947A} and that other processes like e.g. 
shock heating play an important role.
Furthermore, SAMs including ram pressure stripping alone, 
only report negligible changes in galaxy properties like colours and 
star formation rates \citep{2003ApJ...587..500O,2005MNRAS.361..369L}.

Following \citet{1972ApJ...176....1G} we assume that gas is stripped from 
the satellite once the dynamical pressure is able to overcome the 
gravitational force binding the gas to the satellite. In terms of energy 
deposited within the satellite gas this can be approximated by: 
\begin{equation}\label{ram}
  \dot{E}_{ram}=\mu \rho_{hot}v_{\perp}^3 \pi r_h^2.
\end{equation}
Here we take $v_{\perp}$ as the velocity of the satellite perpendicular to its 
disk orientation and assume that the orbital velocity of the satellite is comparable to 
the sound speed $c_{gas}$ of the hot gas. The efficiency of this process for a face-on 
disk should be maximal and minimal for an edge-on disk and we take this into
account by assuming the disk orientation is random with respect to the infall
direction. This is in good agreement with cosmological dark matter 
simulations which show that the spin vector of merging dark matter halos are 
randomly aligned to each other \citep{2006A&A...445..403K}. We assign a random 
angle $\alpha_{\perp}$ between the disk plane and the velocity vector of the satellite 
and calculate $v_{\perp}$ by $c_{gas} \sin \alpha_{\perp}$ \citep{2005MNRAS.361..369L}.
For simplicity we calculate the density $\rho_{hot}$ by taking 
the average density of hot gas within the host halo's virial radius and 
take $r_h$ to be the characteristic half mass radius of the gas within 
the satellite. We  introduce a free parameter $\mu$ which allows 
us to investigate the importance of this process. As we will show below, 
the efficiency
parameter $\mu$ does not influence general galaxy properties significantly. 
Equation \ref{ram} is an upper
limit to the expected ram pressure heating of the gas in the satellite as we 
neglect tidal stripping  and the change in the 
velocity of the satellite while it  orbits through the host halo 
\citep[see e.g.][]{2001ApJ...559..716T}. 

The amount of cold disk gas stripped from the satellite can be  
calculated by:
\begin{equation}
  \frac{dM_{ram,c}}{dt}=-\frac{4}{3}\frac{\dot{E}_{ram,c}}{f_e V_{max,sat}^2}.
\end{equation}
Here we use in Eq. \ref{ram} the observed mean size-mass relation
reported  for disk galaxies from the SDDS \citep{2003MNRAS.343..978S}.
We note that this might be an overestimate as disk galaxies  
tend to be smaller by a factor of up to $~1.5$ at $z=2.5$ 
\citep{2005astro.ph..4225T}. 
\begin{equation}\label{ram2}
  \frac{dM_{ram,h}}{dt}=-\frac{4}{3}\frac{\dot{E}_{ram,h}}{f_e V_{max,sat}^2}.
\end{equation}

Not only cold gas in disks is subject to ram pressure stripping but also 
diffuse hot halo gas \citep[e.g.][]{2006ApJ...644..155M} and we model 
this process in the same way as for the cold gas case with the exception 
that $r_h=r_{vir,sat}/2$ in Eq. \ref{ram} and that the energy needed to heat 
a unit mass to the virial temperature of the host halo is less since the hot
gas in the satellite is already at some temperature $T_{gas}$. The 
hot gas mass stripped from the satellite is

\begin{figure}
    \plotone{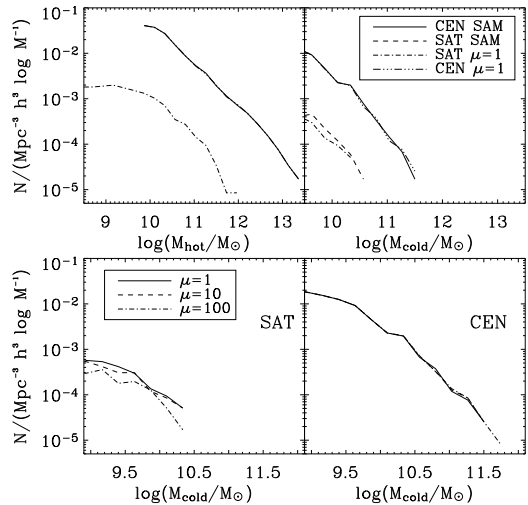} 
    \figcaption{Same as Fig. \ref{fig7}, but in this simulation we only included 
      ram pressure stripping of satellite gas with different stripping 
      efficiencies $\mu$. \label{fig9}}  
\end{figure}

The rate at which material is stripped in our implementation of ram pressure 
varies from earlier approaches. E.g. \citet{2003ApJ...587..500O}  assume
instantaneous stripping of all gas once the ram pressure overcomes 
a characteristic restoring force per unit area in the galactic disk, 
calculated using the surface mass density at the half mass radius. Other 
implementations have a more gradual stripping rate, assuming an exponential profile 
for the surface mass density of material in the satellite disk \citep{2005MNRAS.361..369L}.
Common to both of these implementations is that they assume a radial profile for the gas 
density in the host halo and that the time dependence of the stripping is driven basicly 
by the physical location of the satellite within the halo. In our implementation the time 
dependency is due to the continued transfer of energy to the gaseous disk of the satellite.
For a host halo that is not significantly changing its average gas density with time, 
we deposit a constant amount of energy per unit time 
within the gaseous disk of the satellite. The stripping rate we calculate in this way is
initially larger than that from gradual models like the one of \citet{2005MNRAS.361..369L}. 
Such models however predict an increase in the stripping rate with time, due to the increasing 
ambient gas density while the satellite orbits towards  the centre of the host halo. The 
effect of these different 
implementations on the galaxy populations is very mild. In our model
the cold  and hot gas mass function of central galaxies are mostly 
unaffected by ram pressure stripping  from satellite galaxies in  comparison 
to our initial SAM shown in the upper panel of Fig. \ref{fig9} and in 
agreement with  previous work \citep{2003ApJ...587..500O,2005MNRAS.361..369L}.
The reason we do not find significant changes in the overall galaxy population
is that central galaxies dominate the mass range studied here and that the gas 
fraction of satellites is much less than that of centrals. Increasing the 
efficiency $\mu$ by two orders of magnitude only marginally
reduces the amount of hot and cold gas in satellites. Therefore again we omit using this free 
parameter, i.e. set it to $\mu=1$.

\subsection{Gravitational Heating}\label{grav}
The energy used to expel gas from satellites is contributed from the 
gravitational potential energy that is gained when a satellite becomes bound
in the potential of the primary halo. The majority of dark halos are 
initially on parabolic orbits  with $E_{tot}=0$ \citep{2006A&A...445..403K} before 
they become bound and merge. According to the virial theorem, maximally 
half of the potential energy 
gained could be used to heat the ICM. We here take into account the energy gained by each 
infalling mass unit of gas and subtract from it the energy necessary to expel it 
from the potential of the satellite and to heat it to the virial temperature of the 
host.

The rate at which energy is gained is connected to the rate at which gas is expelled from
satellites $\dot{m}_{gas}$,  according to the prescriptions in section 
\ref{sshock} \& \ref{sram}. At each time step within our simulation we calculate:
\begin{equation}\label{egrav}
 \dot{E}_{grav}=\sum_{i=1,n_{sat}} \dot{m}_{gas,i} \left[ \Delta \phi -b \frac{3}{4}V_{max,sat,i}^2 - \frac{3}{4}V_{max,cen}^2 \right] 
\end{equation}
where $b=2$ for cold gas and $b=f_e$ for hot gas that is stripped from the satellites. 
Please note that Eq. \ref{egrav} is the actual surplus of energy available to heat the ICM 
once the stripped gas of the satellite is heated to the virial temperature of the host. 
We model the dark matter halo of the host as a truncated isothermal sphere  with core 
radius $r_0$ and calculate the   amount of potential energy gained by the satellite when 
it reaches the virial radius $r_{vir,cen}$ of the host halo. For the halo we assume a 
characteristic value $r_{vir,cen}/r_0=25$ \citep{1999MNRAS.307..203S,2002MNRAS.336..119M}. 
It should be noted that in general the concentration of dark matter halos is a function 
of mass and redshift  \citep[e.g.][]{2001MNRAS.321..559B,2004A&A...416..853D} and that 
we ommit this dependency for the sake of simplicity at this stage. The gain in potential 
energy is then calculated as $\Delta \phi= -\ln (r_{vir,cen}/r_0)$.
Within the simulation we will use the energy surplus calculated from  Eq. \ref{egrav} to heat
and to counter cooling within the hot gas of the host halo. 
Eq. \ref{egrav} is generally not included in SAMs, although it appears to be
 necessary in order to conserve energy. 
\begin{figure}
    \plotone{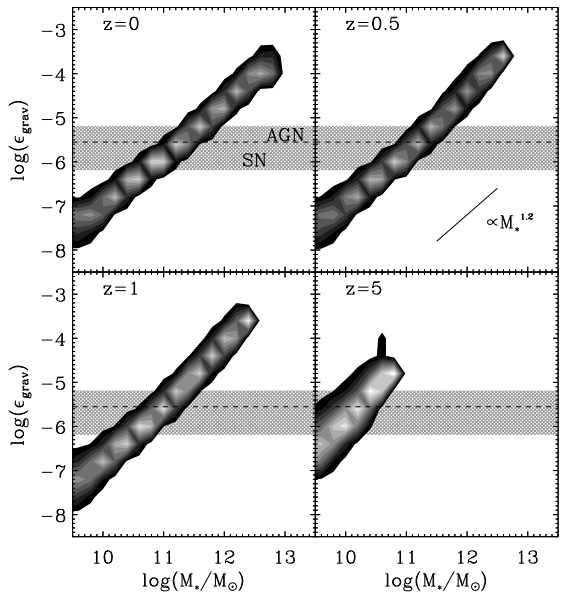} 
    \figcaption{The gravitational heating efficiency $\epsilon_{grav}$ as a 
      function of stellar mass for different redshifts. The crosshatched 
      region shows the region were AGN and supernovae feedback operate. 
      Estimates for supernovae feedback limit it mainly to below the dashed 
      line, while estimates for AGN-feedback cover the whole region. The 
      solid line shows a fit of the form $\propto M_*^{1.2}$.    
      \label{fig14000} } 
\end{figure}

\begin{figure}
    \plotone{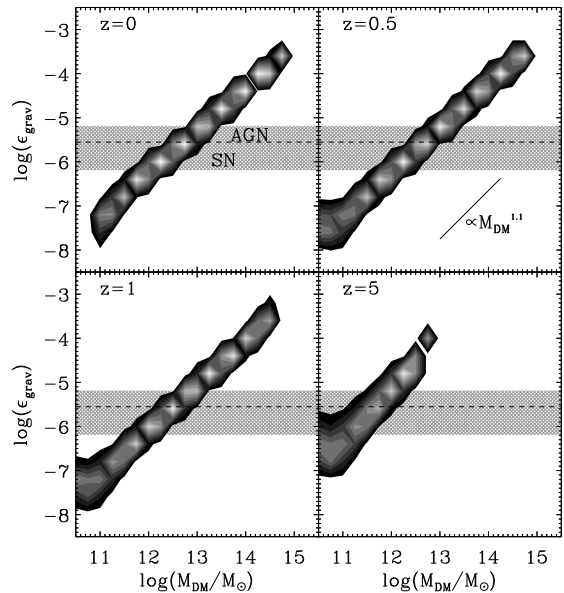} 
    \figcaption{The gravitational heating efficiency $\epsilon_{grav}$ as a 
      function of dark halo mass for different redshifts. Please note that at $z=0$ 
      the contours start at $10^{11}$ M$_{\odot}$  because we sample the mass 
      function only between $10^{11}-10^{15}$ M$_{\odot}$. The crosshatched 
      region and dashed line indicate the regions of 
      AGN and supernovae feedback as in 
      Fig. \ref{fig14000}.  The solid line shows a fit of the form 
      $\propto M_{DM}^{1.1}$. this fit is somewhat steeper 
      than what would be expected from simple scaling arguments and indicates 
      that in massive halos the ratio of expelled satellite gas to stellar mass
      of the central galaxy scales as $ M_{DM}^{0.43}$.\label{fig14001}}  
\end{figure}

The time integral of Eq. \ref{egrav} can be 
quite substantial and, if expressed in terms of $E_{grav,tot}=\epsilon_{grav} m_{*}c^2$,  
we find values in Fig. \ref{fig14000} for $\epsilon_{grav}$ ranging from a few times $10^{-8}$  
to a few times $10^{-4}$ in galaxies of $\sim 10^{10}$ and $\sim 5 \times 10^{12}$ M$_{\odot}$, 
respectively. 
That this increase is driven by the environment becomes most evident when considering 
the dependence of
$\epsilon_{grav}$ on the dark matter halo mass in Fig. \ref{fig14001}. Above a halo mass of 
$10^{11}$ M$_{\odot}$  $\epsilon_{grav}$  increases steadily. However, it appears that this 
increase is steeper at larger redshifts. We find that the upper limit for $\epsilon_{grav}$ 
lies around $\sim 5 \times 10^{-4}$ and that in general the most massive halos present at a given 
redshift tend to take on this values. One can understand this behaviour by considering the accretion 
of satellites onto halos. In general the most massive halos at any redshift have had the largest 
accretion rates in the past which explains the large amount of gravitational heating. 

It is worth 
comparing gravitational heating to other common heating mechanisms like supernovae and AGN. Comparing 
$\epsilon_{grav}$ to $\epsilon_{SN}\sim 2.8 \times 10^{-6}$ and  
$\epsilon_{BH}$ with values between$ \sim 6.5 \times 10^{-6}$ \citep{2005ApJ...620L..79S} and $\sim 6.5 \times 10^{-7}$ \citep{2007astro.ph..3057C}  shows that in general 
gravitational heating is more efficient than supernovae feedback only in galaxies larger 
than a few times 
$ 10^{11}$ M$_{\odot}$ and in halos more massive than $5 \times 10^{12}$ M$_{\odot}$ 
at $z=0$. 
This regime correspond to  massive field galaxies and extends into group like 
environments. For even more massive galxies  and dark halo masses larger than 
$10^{13}$ M$_{\odot}$ gravitational heating starts dominating over proposed AGN-feedback rates. 
This is very interesting as
one is dealing with mostly rich group and cluster environments which are subject to a very 
substantial gravitational heating generally neglected within SAMs. 
Furthermore, the  most massive galaxies 
will  have another large source of  heating in addition to AGN-feedback that will prevent
 ongoing star formation and will naturally reduce the overproduction of galaxies at the 
bright end of the luminosity function \citep{2003ApJ...599...38B}. 
Again it is important to understand at which mass scales gravitational 
feedback operates at different times. At large redshifts it will become more important 
than supernovae and AGN-feedback 
already in smaller halos and galaxies. As a consequence gravitational feedback can match AGN-feedback
 at the  epoch of peak QSO-activity in the most massive halos and galaxies.

When considering the mass dependence of $\epsilon_{grav}$ we find over a wide range of masses roughly a $\propto M_{DM}^{1.1}$ dependency. It is worth noting that from simple scaling arguments one would expect that $\epsilon_{grav} \propto M_{DM}^{2/3} m_{gas}/m_* $, with $m_{gas}$ as the total amount of gas expelled from satellite galaxies, indicating that the fraction of expelled gas to stellar mass of central galaxies scales as $ \propto M_{DM}^{0.43}$.  Going to smaller masses
there is a distinctive break at a stellar mass of a few times $10^{10}$ M$_{\odot}$, which coincides 
with the break in properties of the galaxy population that is found in the Sloan Digital Sky Survey 
\citep{2003MNRAS.341...54K}. Our results suggest that this break could be the consequence of the 
accretion rate and hence the amount of gravitational energy that is released.
\begin{figure}
    \plotone{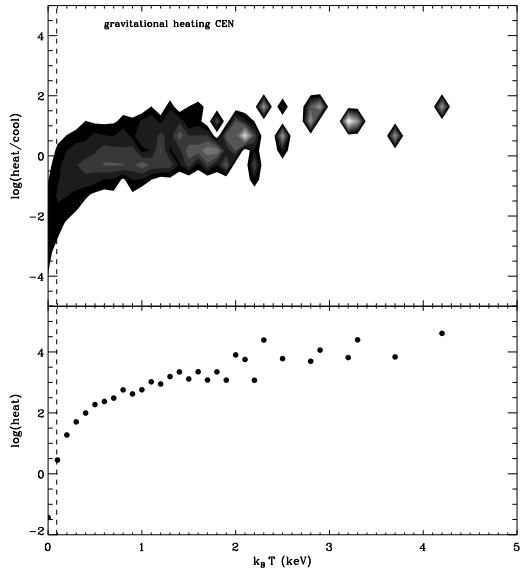} 
    \figcaption{Contours of the effective heating-to-cooling rate 
      $p(\log(heat/cool)|k_{B} T)$ at $z=0.1$ as a function of the environment.       
      We refer with {\it heat} to the amount of cold gas in 
      $M_{\odot}/$yr that can be heated to the virial temperature of 
      a galaxie's dark matter halo by gravitational potential energy 
      released, and with {\it cool} to the amount of radiative 
      cooling in $M_{\odot}/$yr that occurs at the same time. In effect the 
      cold gas reservoir of a galaxy stays constant if 
      $\log(heat/cool)=0$. 
      The top panel shows the heating-to-cooling rate of gas in 
      central galaxies and the small panel on the bottom of each graph shows 
      the absolute amount of heating in units of $M_{\odot}/$yr. 
      The environment is indicated by the virial 
      temperature of the dark matter halo associated with the central galaxy. 
      The vertical dahed line in each graph 
      shows the mass resolution of our simulation. \label{fig14}} 
\end{figure}

\subsection{Implementation}\label{impl}
At the beginning of our merger tree, i.e. when the dark matter halo first crosses $M_{min}$ we 
set the temperature of the hot gas to the virial temperature  $T_{vir}$ of that halo and allow it to
cool as described in \ref{cool}. Once satellites fall  onto the main halo, we
calculate the amount of gas stripped from each individual orbiting satellite $\dot{m}_{gas}$ 
using Eqs. \ref{shock}, \ref{shock2}, \ref{ram} \& \ref{ram2} plus the contribution from gas 
that leaves the satellite halo because of supernovae feedback. 
This $\dot{m}_{gas}$ is then used in Eq. \ref{egrav} to calculate 
the amount of gravitational heating energy added to the ICM. Please not that in practice we use 
Eq. \ref{egrav} without the last term $3/4 V^2_{max,cen}$, because the hot gas of the host might 
not be at $T_{vir}$, but at some lower temperature $T_{gas}$, and therefor the stripped gas will  
not automatically be heated to $T_{vir}$ as assumed in Eq. \ref{egrav}. The energy contribtuion we 
calculated in this way is then used to heat the host hot gas to $T_{vir}$. If the gas is allready
at $T_{vir}$ or if energy is left after elevating it to $T_{vir}$ we allow the surplus of energy
to be used to counter the energy losses due to our cooling prescription in \ref{cool} and 
to reduce the amount of cooling gas. If there is still energy left after countering all the energy 
losses due to cooling we use it to increase the energy of the host hot gas. We here do not take into
account the possiblility of gas leaving the host halo and getting lost once its energy gets 
to large to be bound, but instead  assume it is marginally bound in a hot athmosphere. 
If two host halos merge we assume that the smaller one becomes a satellite and 
calculate $f_e$ according to its specific energy and use it in Eqs. \ref{shock}, 
\ref{shock2}, \ref{ram} \& \ref{ram2}. The satellites within this halo will now be considered 
satellites of the new host halo and contribute their potential energy to the new host.
\begin{figure*}
    \plotone{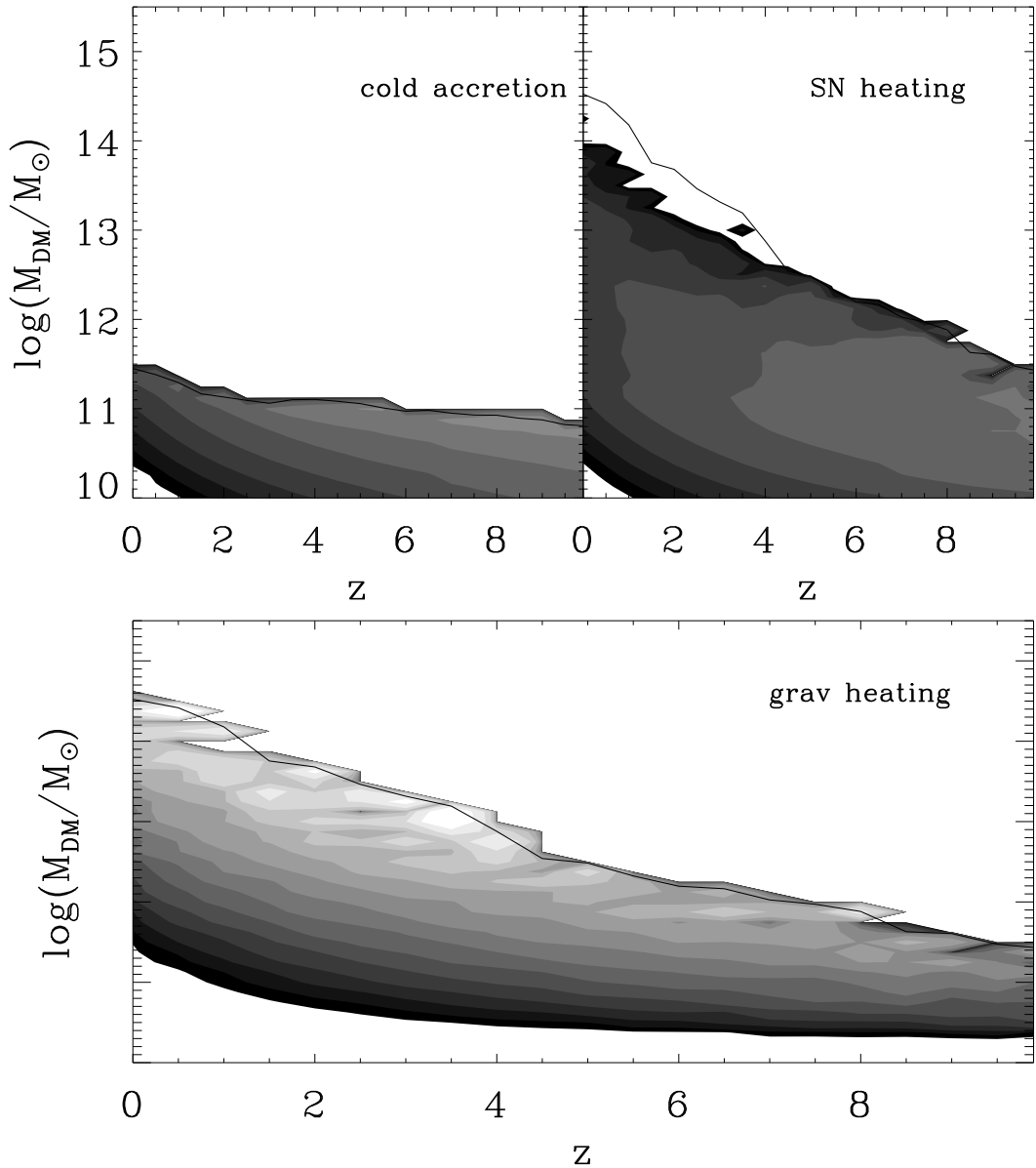} 
    \figcaption{The rate of cooling (heating) as a function of dark halo mass 
      and redshift. The solid  line in the top left panel shows the maximum 
      halo mass in which cold accretion occurs, while the solid line in the 
      other panels shows the maximum dark halo mass we find at any given 
      redshift. Contours have a factor of two difference from each other. The 
      level of the contours starting with the color black and increasing 
      toward lighter colors are $1, 2, 4, 8...8192$ . (\label{fig14b}} 
\end{figure*}

\section{General Results}
To illustrate the contribution from gravitational  heating we display 
the contours of the conditional probability for the ratio of heating to cooling
that individual galaxies experience in a given environment.
The top panel in Fig. \ref{fig14} shows the probability contours for central galaxies  
 at $z=0.1$ in our simulaitons.
 We translate the deposited energy per
 unit time into a heating rate, labeled {\it heat} in Fig. \ref{fig14}, 
by calculating the amount of cold gas that can be heated to the virial 
temperature of the dark matter halo the galaxy resides in. 
The cooling rate for central galaxies is calculated using the prescription outlined in  
section \ref{cool} and is labeled {\it cool} in Fig. \ref{fig14}. 

The left panel shows the contribution from gravitational heating to the 
hot gas of central galaxies as gas gets stripped from satellites. The heating rate
for the central galaxies is up to $10^2$ times larger than the cooling rate and in 
the most dense environments the heating rate becomes 
$~ 10^4$ M$_{\odot}$ yr$^{-1}$. The heating rate shows a clear environmental 
dependence reflecting the higher abundance of satellites which contribute to 
gravitational heating. From these results one expects that star formation will be
terminated in central galaxies of dense environments. 
\begin{figure}
    \plotone{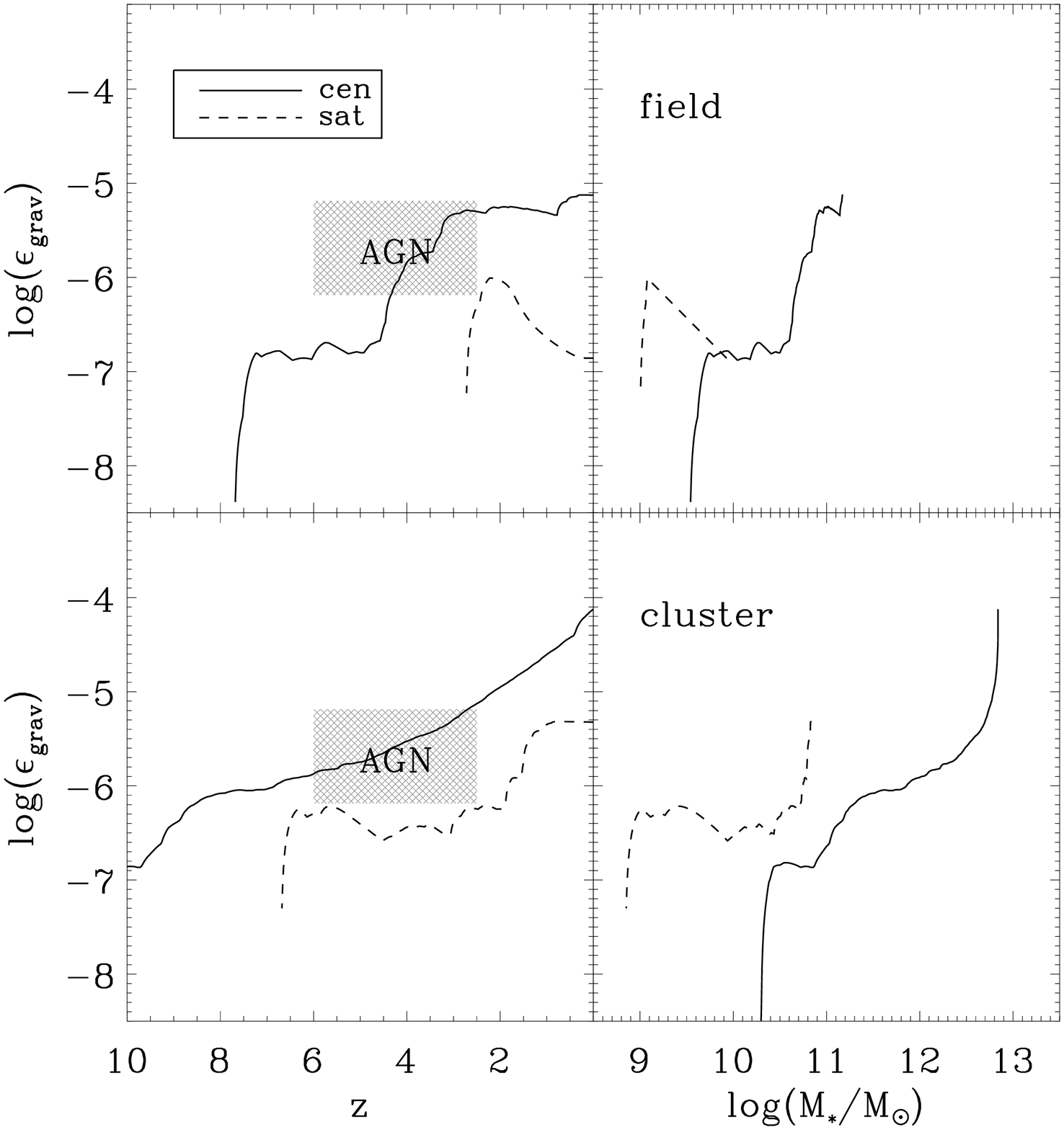} 
    \figcaption{The $\epsilon-$trajectory of gravitational heating 
      in field galaxies (top) and cluster galaxies (bottom).
      We show two set of results for random central galaxies (solid line) and random satellite 
      galaxies (dashed line).  We calculate $\epsilon_{grav}$ by summing it up over all 
      progenitor galaxies. We do the same
      for the masses shown on the x-axes of the panels on the right. The shaded area labeled AGN
      shows the range over which AGN-activity is expected. \label{fig45450}} 
\end{figure}

The importance of several physical processes depending on environment 
and redshift is shown in Fig. \ref{fig14b}. The top left panel shows the 
average cooling rate in systems for which the cooling time is shorter 
than the dynamical time of the halo. This 'cold accretion' mode 
occurs in halos with masses below $\sim 10^{11}-10^{12}$ M$_{\odot}$ and is most 
efficient at high redshifts in agreement with SPH-simulations by \citet{2005MNRAS.363....2K} and earlier analytic calculations \citep{1977ApJ...215..483B,1977ApJ...211..638S,1977MNRAS.179..541R}. 
The solid line in the same figure is the maximum 
halo mass which shows cold accretion at each redshift. The average amount of cold
gas in the ISM reheated by supernovae is shown in the right top panel. Supernovae 
are able to heat gas efficiently in small halos with masses below 
$\sim 10^{12}$ M$_{\odot}$ and are responsible for shaping the low mass tail of 
the luminosity function \citep{1986ApJ...303...39D}. The results in the two top panels are 
usually incorporated in all semi-analytic models. The additional physics we 
included is shown in the lower panel of the same figure. It appears that the 
heating rate of the ICM by gravitational heating 
is always very efficient in the most massive halos one finds at each redshift, 
with  several thousand solar masses per year, and not 
very efficient in low mass halos. This is not too surprising considering that low mass
halos do have  less infalling sub-sructure than cluster sized halos.  

To illustrate how the new environmental effect operates  
we followed  the $\epsilon-$'trajectory'   of a random central and  satellite galaxy back in time.
This trajectory is calculated as in \citet{2006MNRAS.370..902K,2006ApJ...648L..21K} by summing 
up the $\epsilon$ from all the progenitors present at a given redshift. The results for that
are shown  in the left panel of Fig. \ref{fig45450}. 
The mass on the x-axes of the right panels is the sum of the 
stellar mass of all progenitors present at that redshift.
We here distinguish between field and cluster environment by associating the field  
with a dark matter halo of $10^{12}$ M$_{\odot}$ and the cluster with a dark matter halo of 
$10^{15}$ M$_{\odot}$ at $z=0$. 

For field galaxies at early times ($z> 5$), when the mass in progenitors is still 
small ($ < 10^{10}$ M$_{\odot}$), gravitational heating is not important. 
Heating/feedback will be dominantly provided by supernovae and 
some, if any,  AGNs in massive galaxies. At late times gravitational heating is catching up with 
supernovae. Central and Satellite  galaxies in the field are very unlikely to develop 
strong gravitational heating at any time during their evolution, as a 
consequence in first approximation it might be justified to neglect these effects when 
modelling their evolution. However, the situation is dramatically different for cluster galaxies. 
The central galaxies in these environments, that later become first brightest cluster 
galaxies generally have gravitational heating surpassing supernovae feedback at redshifts around 
$z=5$. Present-day cluster member galaxies on the other hand, depending on their mass,  will 
surpass supernovae feedback at later times when their mass is large enough ($>10^{11}$ M$_{\odot}$) 
and stop increasing in $\epsilon$ once they become cluster members.
It is interesting to note that for the first brightest cluster galaxies gravitational heating 
steadily increases and continues to
become very important at late times and that the overall energy released will exceed that of the AGN 
 but only later when the main AGN activity is already over.

\section{Cosmic Star Formation Rate}
The observed average cosmic star formation rate 
\citep{1996ApJ...460L...1L,1996MNRAS.283.1388M} shows a strong decline at 
redshifts less $z \sim 2$ and a modest decline at $z > 3$ 
\citep[e.g][]{2004ApJ...600L.103G}, a trend generally recovered by 
semi-analytic models with varying accuracy 
\citep[e.g.][]{2000MNRAS.319..168C,2004ApJ...600L.103G}. One of the occurring 
problems in these models is 
the steep decline of the star formation rate at low redshifts while 
reproducing the  star formation rates at high redshifts. 
Different approaches have shown some progress in that respect by e.g. including
feedback form super-massive black holes \citep[e.g.][]
{2006MNRAS.370..645B,2006MNRAS.365...11C} which helps reducing star 
formation in early type galaxies at late times or by the shut off of 
star formation in halos above a critical mass \citep{2006MNRAS.370.1651C}.

\begin{figure}
    \plotone{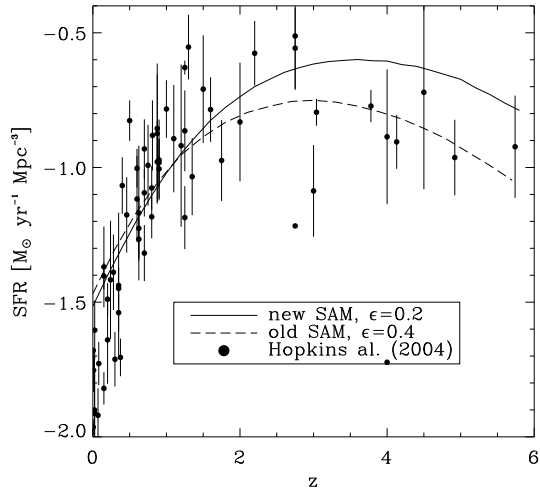} 
    \figcaption{ Modelled vs observed star formation history. We show the best fit
      initial SAM (dashed line) and the best fit new model  
      including all environmental effects (solid line). 
      The observed data is the compilation from \citet{2004ApJ...615..209H} \label{fig15} }
\end{figure}

The environmental effects introduced in the previous sections start operating 
effectively in high density environments like clusters of galaxies and are 
able to prevent cooling of gas and associated star formation. The average 
cosmic star formation rate \citep{1996ApJ...460L...1L,1996MNRAS.283.1388M} 
in that sense provides an ideal way to compare our model predictions and to 
judge the importance of the effects we added. 
Our model prediction, shown as the solid line in  Fig. \ref{fig15}, 
is in  quite good agreement with the observations summarized 
by \citet{2004ApJ...615..209H}.
Our new model differs from the 
best fit initial SAM (dashed line) in one very important point, 
we find  higher star formation
rates at $z > 3$ and significantly lower ones at $z < 2$. Furthermore, we have to 
reduce the energy deposited in the ISM by super-novae as we do have additional 
environmental heating sources that counter the cooling rate and regulate the star formation.
The reason for the change in shape of the cosmic star formation is that the environmental effects
operate not like supernovae feedback which is essentially proportional to the star formation rate, 
but are dependent on the mass of the halo and its 
assembly time. We find that many galaxies which dominate the star formation 
rate at redshifts $z > 4$ and that  end up in dense environments at $z=0$ 
will not yet have assembled in groups and hence  gravitational heating  will not 
be significant. Once the assembly starts taking place during the peak
of the merger epoch around $z\sim 2$ \citep{2001ApJ...561..517K} 
the environmental effects start to operate and to provide
more feedback than the supernovae and as a consequence the star formation rate declines 
steeper than in a model with only supernovae feedback.

\begin{figure}
    \plotone{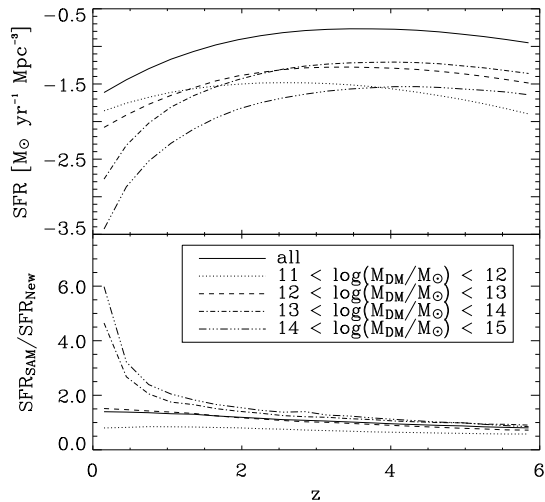} 
    \figcaption{ Top panel: Cosmic star formation history for the model 
      including all environmental effects. We show results for different 
      environments as indicated by the dark matter halo mass, with larger halo masses 
      representing denser environments. Note how star formation at late times is dominated by low mass galaxies in low density environments. 
    Bottom panel: Ratio of star formation rates in different environments 
    between the initial SAM and the model including the environmental
  effects using the same supernovae feedback efficiency. \label{fig15b}} 
\end{figure}

In Fig. \ref{fig15b} we split the contribution to the cosmic star formation rate into
 different environments according to the host dark matter halo mass. The top panel 
illustrates nicely the steep decline of the star formation rate in cluster
 environments compared to field environments, a signature of down-sizing by 
environment. In the lower panel we show how the new environmental prescriptions 
relate to the old SAM prescription with same $\epsilon$. 
As we mentioned above at low redshifts the star formation 
is heavily suppressed by up to a factor of six in cluster environments due to our  
implementation of gravitational heating, while in field 
environments not much changes. Most of the star formation at late times occurs in low mass systems in moderate density environments.

\section{Down-sizing}
\begin{figure}
    \plotone{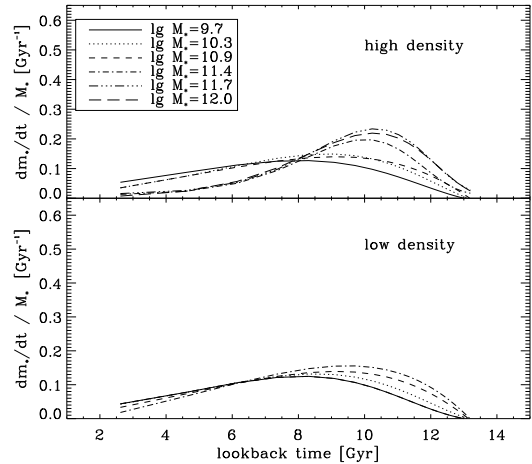} 
    \figcaption{The specific star formation rate in units of Gyr$^{-1}$ for galaxies 
      of different present day masses in the old SAM. The upper panel shows only 
      galaxies in cluster environments with M$_{DM}> 10^{14}$ M$_{\odot}$ and the 
      lower panel only galaxies in field and small group environments with  
      M$_{DM}< 10^{14}$ M$_{\odot}$. The lines are fits to the simulation
      data.\label{down1}}  
\end{figure}

\begin{figure}
    \plotone{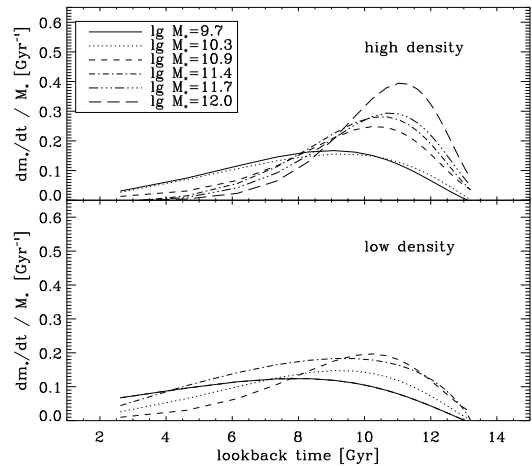} 
    \figcaption{Same as Figure \ref{down1}, but for a SAM including 
      enviormental effects. Note the degree to which low density 
      enviornments dominate at late times.\label{down2}}  
\end{figure}

Growing observational evidence suggests that the main sites of star formation activity
changed  from massive systems at early times to low mass systems 
at late times \citep[e.g.][]{2005ApJ...619L.135J,2005ApJ...621..673T,2007astro.ph..2208Z}. 
A possible explanation for 'down-sizing' from the point of hierarchical modelling is 
that heating overcomes cooling at late times in massive systems 
\citep{2005astro.ph.12235N}. One 
approach to try to address this problem is e.g. using  AGN feedback 
\citep{2005ApJ...635L..13S,2006MNRAS.365...11C,2006MNRAS.370..645B}. We here 
investigate how environmental effects influence down-sizing. As shown in Figs. \ref{fig14001} 
\& \ref{fig14b} environmental  heating is very important for the most massive galaxies and
halos. In addition the mass scale affected by  environmental heating decreases going to 
larger redshifts showing down-sizing.  It is important to note 
that we do not claim that these are the sole responsible effects for causing 
down-sizing but that we investigate their contribution to down-sizing. In Figures
\ref{down1} \& \ref{down2} we show the specific star formation rate $\dot{M}_*/M_*$ 
in units of Gyr$^{-1}$ as a function of the lookback time for galaxies with different
present-day masses. Additonally, we split the sample into galaxies
 residing in massive group/cluster ($M_{DM} > 10^{14}$ M$_{\odot}$) 
and field/small group  ($M_{DM} < 10^{13}$ M$_{\odot}$) environments. 
The results for the best fit old SAM without environmental effects show a long 
extended tail to low redshifts even for the most massive galaxies, and no strong
 difference between massive $\log M_*> 11.4$ and low mass galaxies with 
$\log M_*< 11.4$ (Fig. \ref{down1}). 
In addition there is no strong environmental dependence in 
the specific star formation rate. The new SAM including environmental effects differs
in several fundamental ways which are important with respect to down-sizing. First,
massive galaxies with $\log M_*> 11.4$ have a strongly peaked star formation epoch
 at lookback times around 11 Gyrs with a strong decline to 
 smaller lookback times and low mass galaxies with $\log M_*<  11.4$ have only 
a modest decline showing the same trend as expected from down-sizing. 
Second, we find a strong environmental dependence in form of galaxies more massive than  
$\log M_* \sim  11.4$ being extremely rare in the field and low mass galaxies
showing more star formation at late times in low dense environments. It is 
interesting to note that we do not find that galaxies of the same mass are 
significantly older in high density environments.

\section{Luminosity Function}
\begin{figure}
    \plotone{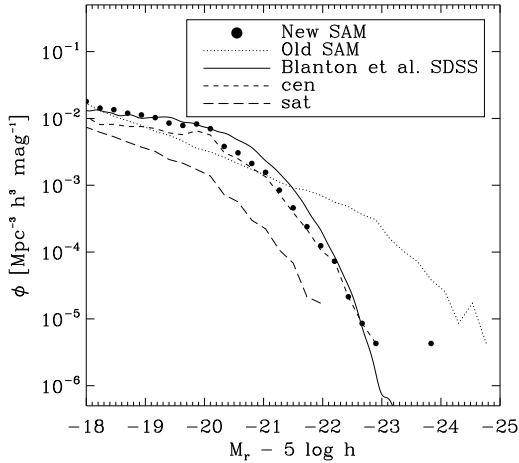} 
    \figcaption{Comparison between the $r-$band luminosity function 
      from the SDSS of \citet{2003ApJ...592..819B} (solid line) and 
      our model predictions (filled black dots). The luminosity function 
      of the central galaxies is shown by the short dashed line and 
      the luminosity function of the satellite galaxies by the long 
      dashed lines. The dotted line shows the predictions of the old SAM
      with no environmental effects. \label{lf} } 
\end{figure}

One attractive point of AGN-feedback is that it helps with the over-cooling 
problem in SAMs and at the same time helps in fitting the luminosity function at the 
bright end by limiting the mass of the most massive galaxies in the simulations. 
We here compare the prediction of our low redshift luminosity function 
with the SDSS $r-$band 
luminosity function  of \citet{2003ApJ...592..819B}. As shown in Fig. \ref{lf} 
the agreement between the  model and the observations is very good over a wide range 
of luminosities. We over-predict the abundance of very luminous galaxies compared to 
the SDSS luminosity function, which might not be a problem considering that  first 
brightest cluster galaxies are not properly covered by the SDSS photometry  nor are 
star burst galaxies (ULIRG) having most of their energy output in the far infra-red part of 
the spectrum. 
Another reason for finding a few too luminous galaxies is that we follow the merging 
of galaxies using a  simplified model based on dynamical friction which is likely
 to overpredict the number of mergers for massive galaxies. It has been shown
 by \citet{spr01} that in these cases the luminosity function in clusters 
shows too many luminous galaxies compared to high resolution simulations
that follow the orbits of galaxies in clusters.

\section{Colour-Bi-Modality}
\begin{figure}
    \plotone{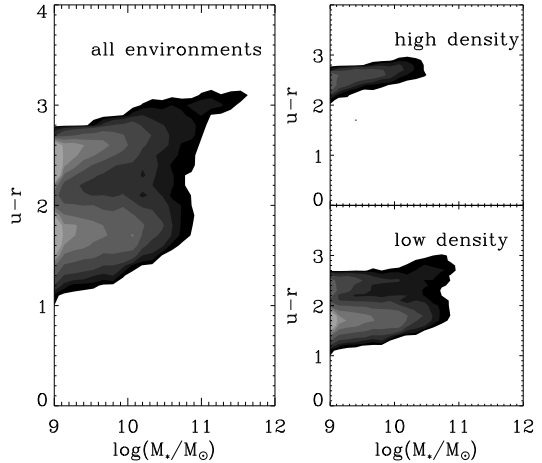} 
    \figcaption{ The colour-mass relation found in our simulations. Left panel shows the overall distribution and the top and bottom right panel the 
colour-magnitude relation in high and low-density environments, respectively, using the same environment definition as in Fig. \ref{down1}.  \label{color}}  
\end{figure}

The results by the SDSS \citep{2006MNRAS.373..469B}
 or surveys like COMBO-17 \citep{2004ApJ...608..752B} show that that the 
galaxy population can be divided according to e.g. its $u-r$-colour into two 
separate regions with a so-called red-sequence of mostly early-type old 
non-star forming galaxies and a blue cloud of mainly late-type star forming galaxies.
Until recently SAMs  have had problems in recovering a strong pronounced bi-modality
and only  with the inclusion of feedback processes or other cooling shut-off
mechanisms could a good agreement to the data be achieved \citep{2006MNRAS.370.1651C,2006MNRAS.365...11C,2006MNRAS.370..645B}. Although these models are 
successful in reproducing the overall distribution of colour in the galaxy population,
they find different galaxies populating their red-sequence and blue cloud. We here 
predict the colour distribution of our model galaxies based on the inclusion of the 
environmental effects. 
The effect of dust on galaxy colors is estimated using the plane-parallel slab model of
\citet{1999MNRAS.303..188K}. For  details of this model we refer the reader to 
\citet{1999MNRAS.303..188K} and to references therein.
 Fig. \ref{color} shows the colour-mass diagram and it is 
very clear to see that we produce a very pronounced bi-modality. The low mass part of
our red sequence is dominated by satellite galaxies being part of a massive group or
cluster environment. This is in agreement with recent findings of \citet{2006ApJ...647L..21H}.
The blue sequence at the low mass end on the other hand is dominated by star forming
central galaxies in field environments. At mass $\log M_*> 10.2$ central galaxies 
start to leave the blue sequence and occupy the red sequence with $u-r> 2.5$. A 
detailed comparison to the data of \citet{2006MNRAS.373..469B} will be presented in a later 
paper in which we combine our SAM with a large-scale N-body simulation.

\section{Discussion and Conclusions}
In this paper we presented a first step in including physical effects that are connected to the 
environment in which galaxies reside into a SAM. Our  approach is novel in that we consider how much 
gravitational potential energy  can be released by gas that is stripped from satellite 
galaxies, once one takes into account the energy needed to strip it. It turns out that this 
source of gravitational heating is very dependent on the environment, as the amount of potential
energy gained for a unit mass coming from infinity increases with the mass of the dark matter 
halo, which is a good proxy for the density of the environment and we find a scaling with
dark matter  mass of $ \propto M_{DM}^{1.1}$.

Gravitational heating in general is more important for galaxies that reside in environments that can
secure a steady infall of  gas-rich satellite galaxies, whose stripped gas contributes
  potential energy. This is naturally the case for present day clusters and massive groups, 
and we indeed find in our simulations most of the gravitational heating occurring in these 
environments. At earlier times the sites of gravitational heating turn to somewhat smaller dark
matter halos and hence less dense environments. This is not surprising considering that the 
mass of $2\sigma$ halos, the mass we can associate with massive groups and clusters today, 
decreases at earlier times. The build up of dark matter on the scales we consider here
is approximately self-similar and one would expect roughly as much substructure falling into a $2\sigma$ halo 
at early times as on a $2\sigma$ halo  at late times. The main difference will be in the 
higher gas fraction of the infalling satellites at earlier times. The higher gas fraction 
almost compensates for the lower potential energy per unit mass in the $2\sigma$ halos at earlier 
times causing them to have similar $\epsilon_{grav}$ as their counter parts at low redshifts.
In this respect our environmental effects are purely driven by the dark matter formation path.

One of the natural outcomes from including gravitational heating is down-sizing 
\citep{2007astro.ph..2208Z} in the star-formation rate of massive galaxies. Central galaxies residing
in the most dense environments generally have a very significant contribution by gravitational 
heating with $\epsilon_{grav} \sim 3 \times 10^{-4}$ which is greater than or comparable
 to the amount 
of feedback from AGNs \citep{2005ApJ...620L..79S,2007astro.ph..3057C}. However, the way that these two heating sources 
operate is very different. While luminous AGNs will heat most efficient during a QSO-phase that takes 
place at  redshifts $z >3$ \citep{2005A&A...441..417H} gravitational heating will start heating
more efficiently than AGNs at redshifts $ z <2$. This is connected to the epoch when most massive environments
assemble by  merging of groups. Besides being effective at late times, gravitational heating has
the additional feature of  showing a strong mass dependence that can  produce the right 
trend of down-sizing in the star formation rate of galaxies. 
Low mass galaxies around $10^{11}$ M$_{\odot}$ generally have 
$\epsilon_{grav} \leq \epsilon_{SN}$  and star formation is regulated mainly by supernovae feedback 
which by itself will not produce down-sizing. Galaxies more massive than that have 
$\epsilon_{grav} > \epsilon_{SN}$  and star formation will be  regulated by gravitational heating.
In summary the following picture emerges, at large redshifts a first episode of strong heating occurs
when AGNs have their peak activity. During that phase a great deal of energy will be deposited within the 
ISM/ICM. After that gravitational heating will start kicking in mainly in the most massive 
systems regulating the cooling rate of gas and hence the star formation rate. We thus argue that 
down-sizing  at low redshifts is a consequence of the redshift and mass dependence of gravitational
heating, which is driven by the environment.

In a study of galaxy properties within the Sloan Digital Sky Survey, \citet{2003MNRAS.341...54K}
find that galaxy properties show a bimodal distribution around 
$M_{crit}=3 \times 10^{10}$ M$_{\odot}$.  Galaxies more massive than this 
have generally older
stellar populations than those less massive. It is an intriguing question to ask what could be 
the origin  for this abrupt transition \citep[see e.g.][]{2006MNRAS.368....2D}. The reason 
for the transition must be closely related the ability to make stars. One natural suspect in this 
respect is feedback. The general prime suspect for feedback, supernovae and AGN feedback, do not 
show a characteristic mass scale at which their action occurs.
The energy output per unit stellar mass is independent of the mass of the galaxy for 
both AGN and SN feedback, but increases as $M_*^{1.2}$ for gravitational heating.
However, it should be noted that the effects of a given amount of heating can be scale dependent and 
hence introduce a characteristic mass scale for supernovae and AGN feedback as well.
Additionally we find that gravitational heating shows a distinctive 
transition  at a mass scale of $\sim 3 \times 10^{10}$ M$_{\odot}$ from being roughly 
constant at smaller masses to increasing steadily at higher masses.  Galaxies below $M_{crit}$ in 
our simulation are mostly unaffected by gravitational heating and hence their star formation is 
regulated by supernovae feedback. Above $M_{crit}$ the situation changes when gravitational 
heating  start becoming more important and able to influence star formation by 
contributing significant amounts of feedback. Even when $\epsilon_{grav} < \epsilon_{SN}$, 
it will be important as a source of feedback and regulator for star formation,
because it operates  independently of the star formation in contrast to supernovae feedback, and 
it will contribute significant amount of heating energy at late times. 
Again, just as in the case of down-sizing,  the feature that could cause the transition mass scale 
$M_{crit}$ is the specific mass dependency and epoch when gravitational heating kicks in.

It is interesting to make the connection between our work and previous work by \citet{1999ApJ...522..590B} who suggest  a scale dependent bias for the galaxy population, in a sense that 
massive,  red galaxies reside within high density environments. Combining this with results 
of  \citet{2006ApJ...653..881N} suggest that one should expect a drop in the star formation 
rate for present-day massive red galaxies that reside in massive environments, 
just as we predicted within our model.

In this paper we included gravitational heating effects in our SAM based on 
simplified physical models and 
made some first predictions/comparisons. The results so far seem very promising and we will
present more detailed comparisons to observations in a follow-up paper. It is clear that this can only
be viewed as a first step in trying to include environmental effects and that further 
comparisons to high resolution hydrodynamical simulations will be necessary to improve 
on this effort.\\

We would like to thank the anonymous referee for his comments and suggestions that helped 
to significantly improve the paper. Also we would like to thank Rachel Somerville, Thorsten Naab, 
Andi Burkert and Joe Silk for usefull comments.

\clearpage

\end{document}